\documentclass[aps,prb,twocolumn,groupedaddress,showpacs,longbibliography]{revtex4-1}

\usepackage{graphicx}
\usepackage{units}
\usepackage{textcomp}

\begin{document}

\preprint{IVNC/APersaud}

\title{Development of a compact neutron source based on field ionization processes}

\author{Arun Persaud}
\email{APersaud@lbl.gov}
\author{Ian Allen}
\author{Michael R. Dickinson}
\author{Thomas Schenkel}
\affiliation{ E.O. Lawrence Berkeley National Laboratory\\
Berkeley, CA 94720, USA}

\author{Rehan Kapadia}
\author{Kuniharu Takei}
\author{Ali Javey}
\affiliation{Department of Electrical Engineering and Computer
  Sciences\\
University of California at Berkeley\\
Berkeley, CA 94720, USA}

\begin{abstract}
  The authors report on the use of carbon
  nanofiber nanoemitters to ionize deuterium atoms for the generation of
  neutrons in a deuterium-deuterium reaction in a preloaded
  target. Acceleration voltages in the range of \unit[50--80]{kV} are
  used. Field emission of electrons is investigated to characterize
  the emitters. The experimental setup and sample preparation are
  described and first data of neutron production are
  presented. Ongoing experiments to increase neutron
  production yields by optimizing the field emitter geometry and
  surface conditions are discussed.
\end{abstract}


\maketitle

\section{Introduction}
In recent years, the use of nanoemitters for neutron production has
been investigated and progress on several approaches such as field
desorption \cite{solano} and field evaporation sources
\cite{reichenbach} has been reported. In addition, there have been investigations into the use of field ionization for
neutron production \cite{Naranjo2005,Fink2009}.

For field ionization, a strong electric field generated by, e.g., a sharp
tip in combination with the Coulomb potential of an atom (or molecule),
provides a tunnel barrier that is small enough to allow an electron
from the atom to tunnel into the tip and thereby ionize the
atom. Field strengths of the order of \unitfrac[20]{V}{nm} are
needed. Field desorption and field evaporation rely on a layer of
adsorbed material or the tip material itself, and require higher
fields than those required for field ionization
($>\unitfrac[40]{V}{nm}$). For comparison, electron field emission requires
fields of the order of \unitfrac[2]{V}{nm}\cite{Gomer1994}.

A device using nanoemitters can achieve a compact and inexpensive
source with a low energy budget compared to rf-plasma or Penning
sources and is therefore a candidate for a portable sealed source that
can be used in oil-well logging or more generally as a replacement for
radioactive sources.  Our approach relies on the use of
nanoemitters that are relatively easy to fabricate.

To generate neutrons, we utilize the D(d,n)$^3$He reaction by
accelerating deuterium ions onto a stationary, deuterated titanium
target. Acceleration voltages of \unit[50-80]{kV} are needed to
achieve neutron yields in the $\unitfrac[10^8]{n}{s}$ range
\cite{Csikai1987}. We also use this voltage to directly apply the
necessary fields to the emitters by placing the target electrode at an
appropriate distance.

\section{Samples}
\begin{figure}[ht]
  \centering
  \includegraphics[width=\linewidth]{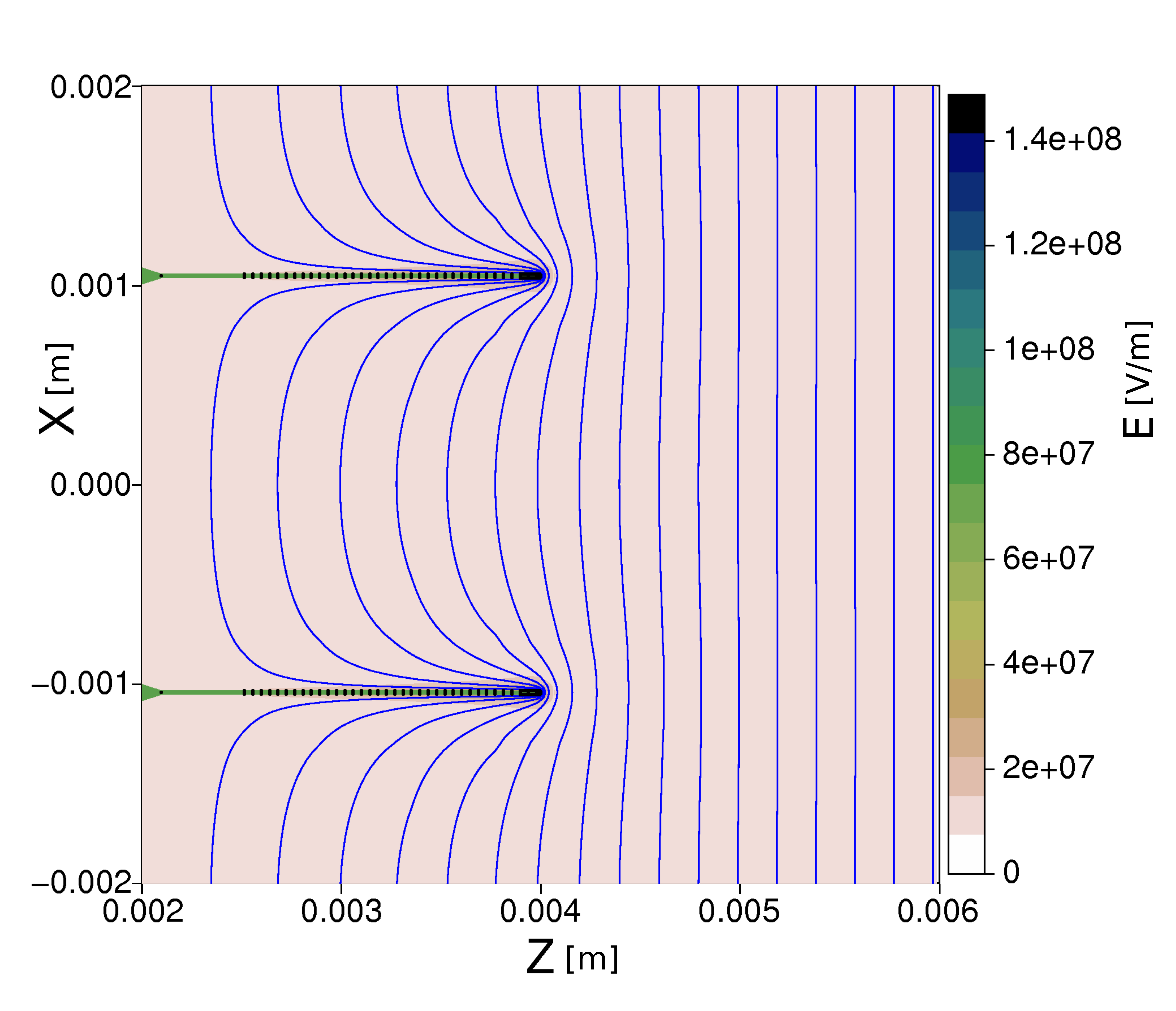}
  \caption{Simulations of field enhancement by an array of high aspect
    ratio tips (arbitrary size). }
  \label{fig:simulations}
\end{figure}

To generate the high fields needed for field ionization, we make use of the
fact that a sharp tip, e.g., a single carbon nanofiber (CNF),
in an electric field compresses the field lines generating
fields that are several thousand times stronger than the field
gradient in a parallel capacitor geometry. For an illustration of the
enhancement effect, see Fig.~\ref{fig:simulations}, which shows the
simulation results of two sharp tips disturbing the field between two
plates. The simulation was done in \texttt{Warp3D}\cite{warp3d}.

Field enhancement allows us to generate the desired field of several
Volts per Angstrom at the tip by applying our acceleration voltage over a
distance of a few centimeters.

\begin{figure}[ht]
  \centering
  \includegraphics[width=\linewidth]{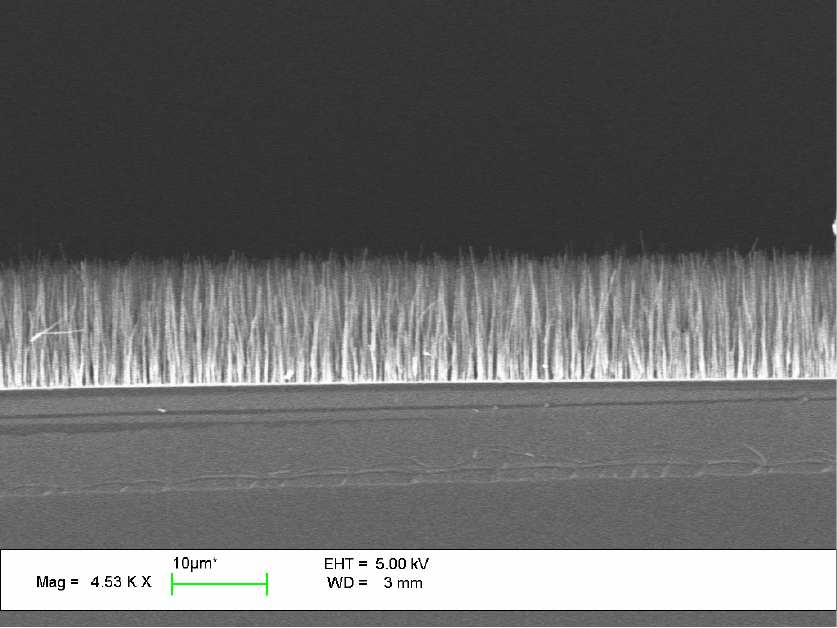}
  \caption{SEM image showing cross section of aligned carbon nanofibers grown on a silicon substrate.}
  \label{fig:forest}
\end{figure}

For the results reported in this article, we use a sample of aligned
multiwall CNFs grown on a silicon wafer in a plasma-enhanced chemical
vapor deposition (PEVCD) process\cite{Meyyappan2003}. A copper layer of
\unit[200]{nm} is deposited on a \unit[50]{nm} oxide layer of a silicon wafer, followed by a \unit[30]{nm} titanium
layer. Then, a final \unit[30]{nm} layer of nickel is deposited. The
nickel layer is used as a catalyst for the CNF growth and the titanium
layer serves as a diffusion barrier for the nickel during the PEVCD
process. The growth process uses NH$_3$ and C$_2$H$_2$ gases in a dc
plasma at \unit[900]{\textcelsius}.  A scanning electron microscope (SEM) image of the CNFs is shown in
Fig.~\ref{fig:forest}. The diameter of the CNFs is about
\unit[70]{nm}. As can be seen in the figure, the growth results in a
relatively uniform height, but single nanofibers also extend above the
forest. These will show the highest enhancement factors.

\section{Setup}
The experimental setup consists of a vacuum system (base pressure
$\unit[10^{-5}]{Pa}$) featuring two electrodes in a parallel plate
arrangement with the possibility to vary the distance between the
plates. One of the electrodes, the target, can be biased to a high
negative voltage (up to \unit[80]{kV}) and features a several
millimeters thick titanium surface that is used to preload the target
with deuterium in a separate process. The other electrode is normally
grounded through a current meter and is used to mount the
nanoemitters. The emitters are exposed to deuterium molecules or
deuterium atoms from a Mantis MGC75 gas cracker source that can
produce a beam of neutral deuterium with a high atomic fraction
($\sim$ 80\%). This is advantageous since accelerated atomic ions will
have a higher kinetic energy per nucleus compared to molecular ions
which translates to a higher D-D reaction cross section yielding a factor of ten higher neutron flux. From Csikai's work\cite{Csikai1987}, we can
expect $\unit[10^5]{n/s/\mu A}$ for \unit[80]{keV} deuterium ions and
$\unit[10^4]{n/s/\mu A}$ for \unit[40]{keV} ions using a fully loaded
target (i.e., Ti$_1$D$_2$).

The neutrons are detected using a Health Physics Instruments Model
6060 neutron detector. We calibrated the detector using AmBe and PuBe
sources of known activity in order to be able to convert from \unitfrac{mRem}{h} to
absolute values of \unitfrac{n}{s}. The background level was
measured over a time frame of several days to \unitfrac[0.5]{counts}{min}.

The setup also allows us to run the emitters in electron emission mode 
by positively biasing the target. We make use of this to characterize the
emitters and to evaluate the field enhancement factor, which can be
obtained by standard Fowler-Nordheim analysis.

\section{Results}
\begin{figure}[ht]
  \centering
  \includegraphics[width=\linewidth]{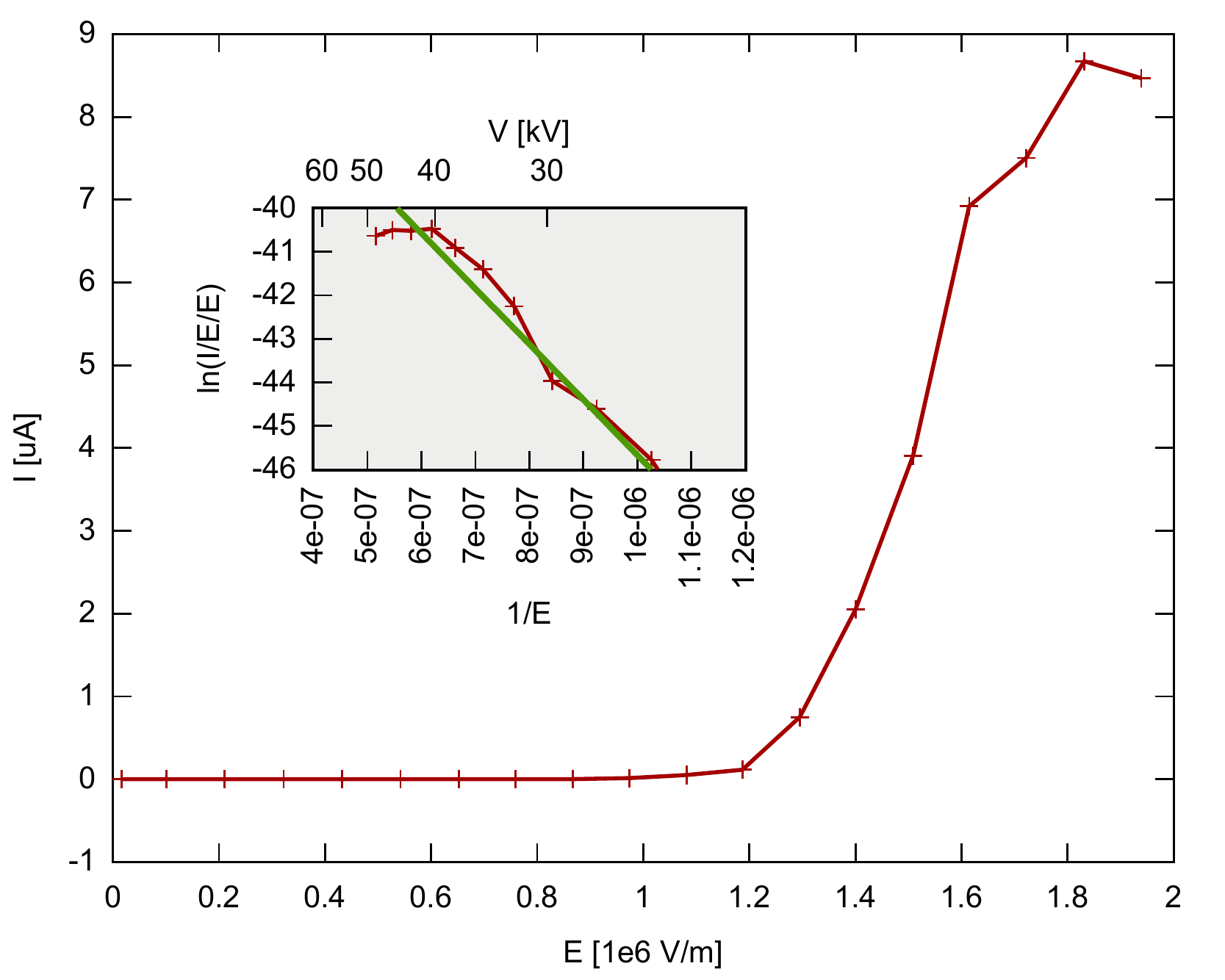}
  \caption{Electron field emission current vs. applied field. Insert shows Fowler-Nordheim plot.}
  \label{fig:FE}
\end{figure}

Field emission studies at a emitter-target distance of \unit[25.4]{mm} and a
pressure of $\unit[10^{-5}]{Pa}$ show onset fields for electron emission in the
range of $\unitfrac[10^6]{V}{m}$, see Fig.~\ref{fig:FE}. The insert in
Fig.~\ref{fig:FE} shows a Fowler-Nordheim plot that is a plot of the
inverse electric field versus $\log(I/E^2)$. For field emission, a
linear relation between these two quantities as in
Eq.~(\ref{eq:FN}) is predicted\cite{Fowler1928} and fits our data well.
The Fowler-Nordheim equation used to fit the data is
\begin{equation}
  \label{eq:FN}
  \log\left(\frac{I}{E^2}\right) = \log\left(\frac{c_aA\gamma^2}{\Phi}\right)-\frac{c_b\Phi^{3/2}}{\gamma}\frac{1}{E},
\end{equation}
where $I$(A) the emitted current, $E$(V/m) the applied field, $\gamma$
the field enhancement factor (so that $\gamma E$ is the local field at
the tip), $A$(m$^2$) the area of the emitter, $\Phi$(eV) the
work function of the material (\unit[4.8]{eV} was assumed for CNF),
$c_a=\unit[1.5414\times10^{-6}]{A \cdot eV\cdot V^{-2}}$ and
$c_b=\unit[6.8309\times10^9]{V\cdot m\cdot eV^{3/2}}$.  From the slope
of the linear fit, the field enhancement factor can be calculated. Our
measurements give an enhancement factor of around $5000$.

\begin{figure}[ht]
  \centering
  \includegraphics[width=\linewidth]{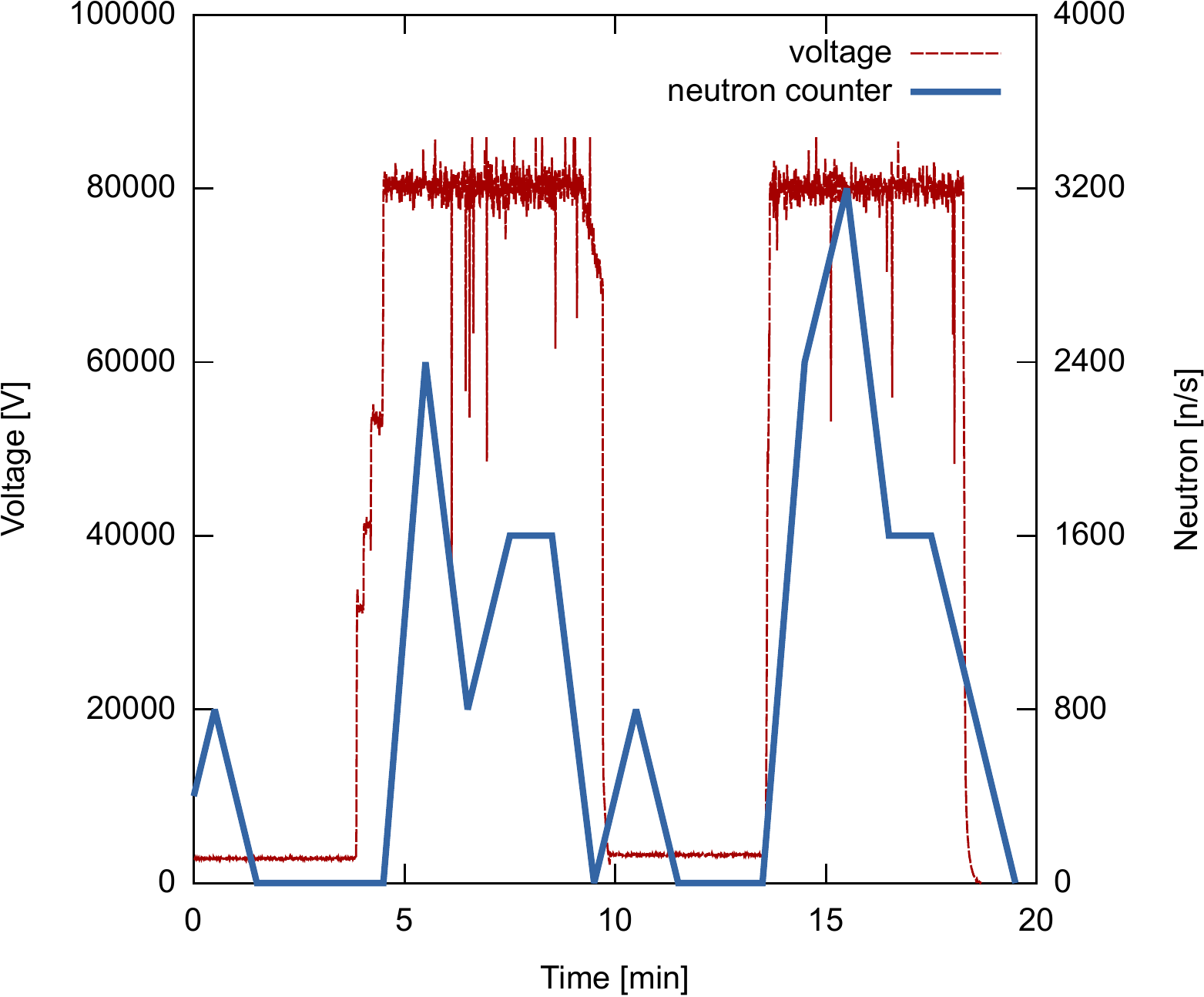}
  \caption{Neutron yields (at the target) from the CNF sample at a
    deuterium gas pressure of
    \unit[0.08]{Pa} and a gas cracker power of \unit[60]{W}.}
  \label{fig:neutrons}
\end{figure}

In Fig.~\ref{fig:neutrons}, first results using CNFs to create neutrons
are shown. The neutrons are generated at an acceleration voltage of
\unit[80]{keV} and a emitter-target distance of \unit[12.7]{mm}. The
gas flow in the chamber was \unit[10]{sccm}, resulting in a pressure
of \unit[0.08]{Pa}. Although the actual count rate at the detector
is small (several counts/minute), a clear correlation between the
applied high voltage and the neutron count rate can be seen and the
signal is clearly above the background level. The
measured neutron rate at the neutron detector corresponds to
\unitfrac[2000-3000]{n}{s} at the target, which agrees with the
measured current in the low microamp range when taking into account
that the target was not fully loaded with deuterium (elastic recoil
detection analysis shows that our samples have a
deuterium-titanium content of about 1:9 in the first \unit[600]{nm} of the
target). The achieved field is also roughly ten times higher than that required for
electron emission, in agreement with values quoted in
literature. Furthermore, neutron production from molecular deuterium
(D$_2$) at these energies produces a neutron yield that is lower by a factor of ten at a current level of $\unit[1]{\mu A}$ which would be below our detection limit.

From the Fowler-Nordheim plot for electron emission, we can also
estimate that only $10^4$ tips contribute to the emission
current. Assuming the same value for field ionization, this translates
into an average current of \unit[100]{pA} per tip which is in the
range of what we expect comparing to \unit[5]{nA} currents that can be
achieved with highly optimized single tips\cite{Naranjo2005} and the
gas pressures used. The process of field desorption requires a higher
field than field ionization and is not expected to contribute at the local fields we achieved.

Considering the density of CNFs of $\unit[2\times10^9]{cm^{-2}}$ and a
coated area of \unit[2]{cm$^2$}, only a very low fraction ($10^{-5}$) of
tips contribute. This is a concern and we will discuss a possible
solution to improve this number in the next section.

High voltage breakdown occured occasionally during the experiment, but
did not degrade the performance of the samples. Inspecting the samples in an
SEM afterwards showed only localized
damage  from sparks (damaged spot size around $\unit[30]{\mu m}$).

\section{Outlook}

\begin{figure}[ht]
  \centering 
  \includegraphics[width=\linewidth]{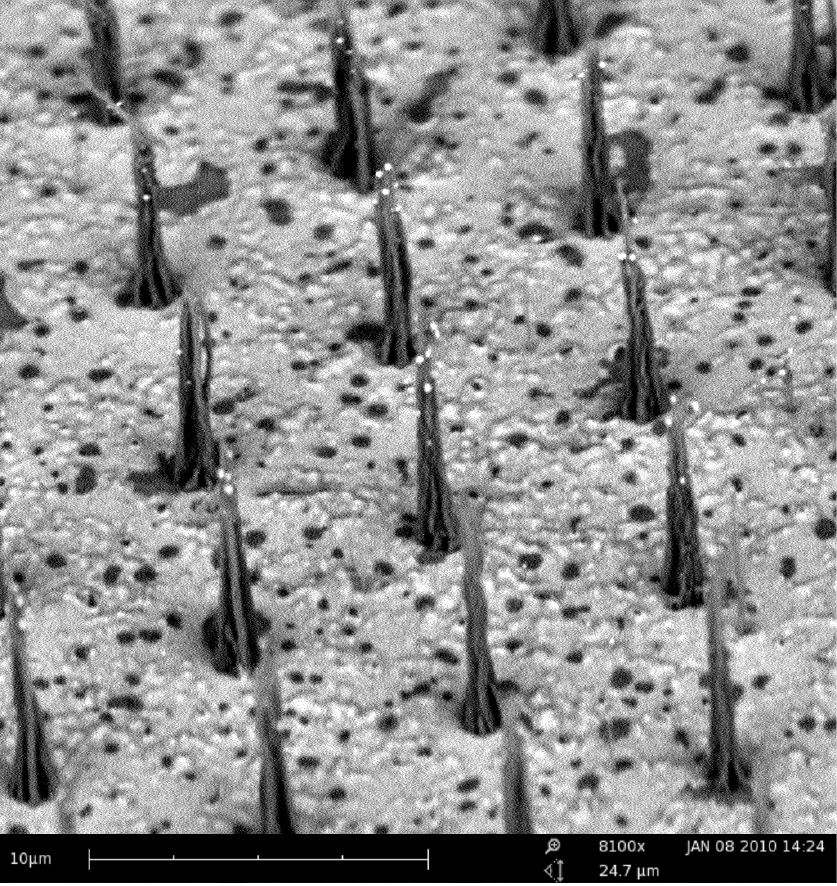} 
  \caption{Array of CNFs with a \unit[6]{$\mu$m} spacing using the
    same growth mechanism (similar growth rate and diameter) as for
    the CNF forest sample.}
  \label{fig:array}
\end{figure}

In ongoing experiments, we are investigating arrays of CNFs such as those
shown in Fig.~\ref{fig:array}. These have the advantage of having a
good separation between tips so that the field enhancement factor of
a single tip is not influenced by emitters that are in close
proximity and therefore a higher fraction of tips is expected to
contribute to the current. In the fabrication process of these samples,
one lithographic step is involved to form the array by patterning the
nickel deposition. The distance between tips can be optimized and it
has been found that a good value for tip
height to tip separation distance is 1:2 \cite{Bonard2001,Jo2003}.

We are in the process of characterizing these new samples and have obtained
electron emission results with enhancement
factors of around $2000$. A possible explanation for the smaller
enhancement factor is that in these arrays the field is shared by several tips. 

Another direction we are investigating is surface-coating of field
emitters. We envision twofold benefits: increased lifetime of the CNFs
due to protection of the structure from hydrogen etching and better field
emitting properties due to optimization of the work function of the
emitting material. The use of different
emitter materials, e.g., etched silicon tips\cite{Rangelow2001},
is another option we are pursuing.

\section{Conclusion}

We show first results from a field ionization neutron generator. The
measured field enhancement factors are comparable with those achieved using single
CNFs. From this fact and the number of contributing tips calculated by
Fowler-Nordheim analysis, we conclude that single CNFs that protrude from the
forest of CNFs (as can be seen in the SEM images) are responsible for
the field ionization current. 

Previous results by Naranjo \textit{et al.}\cite{Naranjo2005} show
currents of \unit[5]{nA} for a single very well conditioned and
optimized tip and they report a neutron yield of \unitfrac[800]{n}{s}
at \unit[115]{kV}. An array of tips with a spacing appropriate to achieve high field enhancement factors should allow for $10^6$
tips per square centimeter, which for optimized tips should allow
neutron yields in the \unitfrac[$10^7-10^8$]{n}{s} range.
The few microamperes of deuterium current reported here are already
sufficient for the generation of $\unitfrac[2\times10^7]{n}{s}$ in a
D-T reaction at \unit[80]{keV} using a fully loaded target, meeting
the yield of currently used AmBe neutron sources. 

For future experiments, we therefore plan to use arrays of single
isolated field emitters to increase the amount of emitting tips while
still generating high field enhancement factors. Furthermore, by
optimizing surface conditions, we plan on increasing field ionization
currents and neutron yields. We expect to be able to reach 
neutron yields suitable for radioactive source replacement or
applications in oil-well logging.

\section*{Acknowledgments}
The authors would like to thank Jeff Bramble and Dave Rodgers for their
help with calibrating the neutron detector, Dave Grothe for help with
the simulations, and Kin Man Yu for the ERDA measurement.

This work was performed under the auspices of the US Department of Energy, NNSA
Office of Nonproliferation Research and Engineering (NA-22) by
Lawrence Berkeley National Laboratory under Contract DE-AC02-05CH11231.


%

\end{document}